\documentclass[manuscript,authorversion,nonacm]{acmart}
\AtBeginDocument{%
  }

\begin{document}

\title[Community-based Approaches to Adolescents Online Safety]{Moving Beyond Parental Control toward Community-based Approaches to Adolescent Online Safety} 
% \title[Engaging Families  Teens in Families' ]{Empowering Teens and their Families to Co-Manage The Privacy and Security Issues of Mobile Technologies}

\author{Mamtaj Akter}
\email{Mamtaj.Akter@nyit.edu}
\orcid{0000-0002-5692-9252}
\affiliation{%
  \institution{New York Institute of Technology}
  \city{New York}
  \state{New York}
  \postcode{10023}
  \country{USA}
}
\author{Jinkyung Katie Park}
\email{Jinkyup@clemson.edu}
\orcid{0000-0002-0804-832X}
\affiliation{%
  \institution{Clemson University}
  \city{Clemson}
  \state{South Carolina}
  \postcode{29634}
  \country{USA}
}

\author{Pamela J. Wisniewski}
\email{Pamwis@stirlab.org}
\orcid{0000-0002-6223-1029}
\affiliation{%
  \institution{Socio-Techincal Interaction Research Lab}
    % \city{Orlando}
  % \state{Florida}
  \postcode{32816}
   \country{USA}
}
% Jinkyung Katie Park, Clemson University, Jinkyup@clemson.edu
% Pamela J. Wisniewski, Socio-Techincal Interaction Research Lab, Pamwis@stirlab.org
\renewcommand{\shortauthors}{Akter et al.}

\begin{abstract}
  % In this position paper, we discuss the paradigm shift that moves away from parental mediation approaches toward collaborative approaches to promote adolescents' online safety. We present empirical studies that explore teen-parent collaborative frameworks to manage online privacy. We also present empirical studies that include a broader community to promote the online privacy and security of teens.

In this position paper, we discuss the paradigm shift that moves away from parental mediation approaches toward collaborative approaches to promote adolescents' online safety. We present empirical studies that highlight the limitations of traditional parental control models and advocate for collaborative, community-driven solutions that prioritize teen empowerment. Specifically, we explore how extending oversight beyond the immediate family to include trusted community members can provide crucial support for teens in managing their online lives. We discuss the potential benefits and challenges of this expanded approach, emphasizing the importance of granular privacy controls and reciprocal support within these networks. Finally, we pose open questions for the research community to consider during the workshop, focusing on the design of "teen-centered" online safety solutions that foster autonomy, awareness, and self-regulation.
\end{abstract}




\maketitle

\section{Introduction}
Today's teens are deeply immersed in the digital world. A Pew Research report indicates that 97\% of U.S. teens use the internet daily, with nearly half (46\%) online almost constantly \cite{vogels_2022_pewresearch}. While social media offers valuable benefits like social connection, creative expression, and peer support \cite{Anderson_2022_pewresearch}, it also exposes teens to a range of online risks. These include cyberbullying \cite{kim2021human, alluhidan2024teen}, exposure to explicit content \cite{park2023towards}, problematic internet use linked to mental health challenges (e.g., suicide contagion) \cite{park2023affordances, tanni_lgbtq_2024}, and even threats to physical safety \cite{valkenburg2022social}. In response to these risks, an overemphasis on restrictive and authoritative parental mediation, such as the use of surveillance-based parental control applications, has become prevalent \cite{modecki2022}.

However, such restrictive approaches often come at a cost. A crucial developmental task of adolescence is the safe and successful transition towards independence and autonomy \cite{park2024s}.  While parental monitoring and control may aim to shield teens from online dangers \cite{alluhidan2024teen}, they can also undermine teen autonomy \cite{baumrind2005patterns}, erode trust between parents and teens \cite{williams2003adolescents}, and negatively impact overall family dynamics \cite{wisniewski2017parental}. Recognizing the shortcomings of purely restrictive strategies, researchers have increasingly explored alternative approaches that prioritize teen empowerment and digital resilience.  One promising direction involves collaborative technologies that move beyond surveillance and instead engage both teens and parents in establishing digital ground rules and jointly managing online activities \cite{akter2022parental, ghosh2018safety, nouwen2017parental, ko2015familync}.

This position paper examines empirical studies that explore this shift from parental control to collaborative parent-teen partnerships for establishing healthy online boundaries and jointly managing online safety, privacy, and security.  Furthermore, we extend this collaborative model beyond the immediate family \cite{akter2024towards} to include broader community support networks (e.g., extended family, friends, neighbors, and colleagues) \cite{akter_it_2023, akter_examining_2025, akter_evaluating_2023, akter2024familydesign}. Finally, we pose open questions for the research community to consider during the workshop. Our position paper is highly relevant to the CHI 2025 Mobile Technology and Teens workshop as our approaches to adolescent online safety are one of the major themes of the workshop (i.e., engaging broader stakeholders and navigating the implications of emerging technologies). 

\section{Moving Away from Parental Control to a Teen-Parent Collaborative Approach}

Research on adolescent online safety has increasingly highlighted the limitations of traditional parental control applications—tools that parents often use to monitor and restrict their children’s online activities. Studies suggest that these applications may not only fail to achieve their intended protective effects but may also harm parent-teen relationships \cite{ali_betrayed_2020, ghosh_understanding_2016, wisniewski_parental_2017, ghosh_safety_2018, moser_parents_2017, pain_paranoid_2006}. For instance, Pain \cite{pain_paranoid_2006} found that excessive surveillance through these apps fosters paranoia and fear among teenagers, ultimately straining trust between parents and teens. 
Furthermore, our previous research \cite{wisniewski_parental_2017} analyzing 75 commercially available parental control apps revealed that many were overly restrictive and invasive, prioritizing parental authority over teen autonomy and open communication. 

In response, scholars advocate for a more balanced approach that fosters transparency and cooperation. 
%Active media mediation (e.g., having parent-teen conversations about media) requires more parenting attention and may be more sensitive to other features of the parent-child relational dynamic. For example, parent-teen connectedness is likely a critical component of effective active media mediation. Active media mediation, in conjunction with a scaffolded approach to restrictive media mediation, allows for greater adolescent autonomy and the development of important skills within digital environments. Active media parenting has also been associated with reduced media-related risks including aggression, substance use, and sleep deprivation. A meta-analysis reported that both restrictive and active media parenting were associated with reduced time online. 
A line of research focused on parent-teen collaboration on the importance of maintaining adolescent privacy and agency in digital spaces. %Wang et al. \cite{wang2021a} propose that both parents and teens prefer applications that promote mutual awareness and feedback, while Cranor et al. \cite{cranor_parents_2014} emphasize the importance of maintaining adolescent privacy and agency in digital spaces.
For instance, Akter et al. developed the "Co-oPS" mobile application \cite{akter_CO-oPS_2022}, designed to provide equal oversight capabilities to both parents and teens. Unlike conventional parental control apps, Co-oPS enables both parties to review each other’s installed applications and privacy permissions, leveraging teens’ technological fluency to support their parents’ data privacy management. To ensure personal privacy, both parents and teens could selectively conceal specific apps from review while allowing them to provide feedback on each other’s visible apps and their permission settings. To evaluate its feasibility, we conducted a lab-based study \cite{akter_from_2022} involving 19 parent-teen pairs. The study assessed their current strategies for managing online safety and privacy, as well as their perceptions of the Co-oPS app. Our findings revealed that both parents and teens often overlooked mobile safety and privacy considerations when installing new apps. While most parents relied on either parental control apps or manual monitoring of their teens’ app usage, teens demonstrated little interest in their parents’ mobile security and privacy. Although both groups valued the transparency provided by Co-oPS regarding app usage and permission settings, they expressed concerns about the CO-oPS’s privacy feature, which allowed users to hide apps from one another. Some of our participants feared that this function could undermine trust within the family. Additionally, power dynamics played a critical role — when using different features of the CO-oPS app, parents were more comfortable reviewing their teens’ app usage and providing feedback, whereas teens showed reluctance in reviewing their parents’ apps and permissions.

\section{From Parent-Teen Collaboration to Community-based Approach}
More recent studies posit that social influences (e.g., peers, family, community) on youth can have a positive impact on adolescents’ motivation and self-regulation. Therefore, taking into account socio-ecological factors in digital parenting can be effective in promoting the digital well-being of adolescent while supporting their autonomy development \cite{park_towards_2023, badillo2020towards}. 
In our prior work, we explored whether the parent-teen collaborative model could be extended beyond the parent-teen dyadic relationship to broader familial and community networks. In the above lab study, parents and teens were also asked to share their perspectives on integrating additional family members into the Co-oPS app to expand oversight within their extended family networks \cite{akter_it_2023}. Both groups saw value in involving the relatives, such as grandparents, siblings, and cousins whom they trusted and cared for, believing that a collective approach could alleviate the burden on parents while ensuring teens' digital safety. Additionally, both teens recognized the potential benefits of including family members who are more vulnerable to mobile privacy threats, such as young children and older adults, allowing for reciprocal guidance and support. However, some participants identified potential challenges associated with this expansion. Parents expressed concerns that extended family members might blame them for their teens' digital behaviors, such as risky app usage or disregarding others' advice. Teens, on the other hand, feared that authoritarian family members might impose additional restrictions on their digital autonomy or engage in intrusive questioning. To address these concerns, we recommended that future app designs incorporate granular privacy controls, allowing users to selectively share app usage information with specific family members.

To further explore the effectiveness of the collaborative approach to mobile online safety, security, and privacy, we recently conducted a field study examining how family members, including teens, collaborate with their broader trusted communities, not just extended relatives, such as friends, neighbors, and co-workers \cite{akter_examining_2025}. From several prior studies, we learned that individuals with greater technological expertise often act as “caregivers,” providing support and guidance to those in the community who are less tech-savvy  ("caregivees") \cite{van_parys_you_2019, correa_brokering_2015, kiesler_troubles_2000, poole_computer_2009, kropczynski_towards_2021}. 
% Conversely, those receiving privacy and security support—often referred to as “caregivees”—tend to rely on others for assistance \cite{kropczynski_towards_2021, anaraky_disclose_2021}. 
We examined the distinct characteristics of these roles within communities and evaluated whether community-based approaches, such as Co-oPS, that treat individuals equally can address such caregiving disparities within communities. In our study, participants self-formed groups within their trusted networks and used Co-oPS for one month, engaging in collaborative digital safety and privacy management. Interestingly, we found that adolescent participants primarily assumed the role of caregivees, while adult participants acted as caregivers. This finding aligns with our other research \cite{akter_from_2022}, which indicated that teens generally exhibit low interest in providing digital oversight for their parents. Furthermore, the study revealed that caregivees benefited significantly more from community-based oversight than caregivers. These results suggest that while teens may not take the initiative in promoting online safety, structured community-based interventions can effectively support their digital well-being. Overall, our findings underscore the potential for extending collaborative oversight beyond the family unit, fostering a broader, community-driven approach to adolescent digital safety and security.

% In sum, our study explored the idea of parent-teens and immediate-extended families' collaboration to oversee one another's online safety and privacy, finding that increased transparency could enhance discussion and mutual learning, ultimately improving family online safety and digital privacy. However, achieving these benefits relies on all family members accepting shared responsibility for each other's online safety and mobile privacy, a paradigm shift from individualistic approach or traditional parental control. 

\section{Future Directions: Designing for Teen Empowerment}
This position paper has presented an overview of empirical studies exploring collaborative approaches to promoting online safety, privacy, and security for teens. Our research strongly suggests a shift away from restrictive parental control models toward community-driven solutions that prioritize teen empowerment. Moving forward, we advocate for the development of online safety tools that empower teens through enhanced autonomy and control.  Crucially, future designs should prioritize features enabling teens to selectively share information with trusted individuals, fostering trust and encouraging their active participation in collaborative oversight.  This approach recognizes teens' growing digital fluency and their need for agency in managing their online lives. Several key questions emerged from our prior research that we aim to explore further during the workshop:

\begin{itemize} 
\item How do socio-ecological factors (e.g., cultural norms, family context) influence adolescent technology use and their experiences with online risks?

\item What are developmentally appropriate, longitudinal approaches to collaborative online safety for teens, and how do these approaches vary based on contextual characteristics (e.g., social, cultural, family context)?

\item Beyond parents and their trusted community, who are the key stakeholders in promoting adolescent online safety? How can we effectively engage them in a broader conversation to support teens?
\end{itemize}
These questions converge on a central theme that we hope to address during the workshop:  \textit{How can we design truly "teen-centered" online safety solutions that empower adolescents by fostering autonomy, promoting awareness, and cultivating self-regulation?}

Participation in the CHI 2025 Mobile Technology and Teens workshop will offer us an invaluable opportunity to connect with researchers dedicated to promoting healthy technology and media use among teens.  We are eager to share our ongoing projects and future research agenda, receive constructive feedback from workshop participants, and explore potential collaborations.  Furthermore, we anticipate gaining valuable insights from the organizers' and other participants' research on teen online safety, enriching our understanding of this critical area.
\section{Acknowledgments}
% \begin{acks}ACKNOWLEDGMENTS
This research was supported by the U.S. National Science Foundation under grants CNS-1814068, CNS-1814110, and CNS-2326901. Any opinions, findings, and conclusions, or recommendations expressed in this material are those of the authors and do not necessarily reflect the views of the research sponsors.

\begin{quote}
\end{quote}
\textbf{Author Bios}\\
\textbf{Mamtaj Akter} is an Assistant Professor at the New York Institute of Technology. Her current research interests include Human-computer Interaction, Adolescent Online Safety, and Usable Privacy and Security. 

\noindent
\textbf{Jinkyung Katie Park} is an Assistant Professor in the School of Computing at Clemson University. Her research focuses on Human-Computer Interaction to promote the online safety of vulnerable populations. 

\noindent
\textbf{Pamela Wisniewski} is a Director of the Socio-Techical Interaction Research Lab. 
Her work lies at the intersection of Human-Computer Interaction, Social Computing, and Privacy.

\bibliographystyle{ACM-Reference-Format}
\bibliography{REFERENCES}

\end{document}